\documentclass[aps,prl,preprint,showpacs,nofootinbib]{revtex4-1}
\usepackage{graphicx}
\usepackage{amsmath}
\usepackage{textcomp}

\begin{document}


\title{Neutrino oscillations and electron-capture storage-ring experiments}

\author{Walter Potzel}
\email{wpotzel@ph.tum.de}

\affiliation{Physik-Department E15, Technische Universit\"at M\"unchen, James-Franck-Str. 1, 85748 Garching, Germany}








\date{\today}

\begin{abstract}
Oscillations in the electron-capture (EC) decay rate observed in storage-ring experiments are reconsidered in connection with the neutrino mass difference. Taking into account that - according to Relativity Theory - time is slowed down in the reference frame of the orbiting charged particles as compared to the neutral particles (neutrinos) moving on a rectilinear path after the EC decay, we derive a value of $\Delta m^{2}_{21}=(0.768\pm0.012)\cdot10^{-4} eV^{2}$ for the neutrino mass-squared difference which fully agrees with that observed in other neutrino-oscillation experiments. To further check the connection between EC-decay oscillations and $\Delta m^{2}_{21}$ we suggest experiments with different orbital speeds, i.e., different values of the Lorentz factor.
\end{abstract}


\maketitle

\subsection{1. Introduction}
Several decay studies with highly-ionized nuclides have been performed at the GSI Helmholtzzentrum f\"ur Schwer\-ionenforschung in Darmstadt \cite{Litvinov2008}, \cite{Kienle2013}. In the most recent and refined experiment hydrogen-like $^{142}_{61}Pm^{60+}$ ions coasting at a velocity $v$ (corresponding to a Lorentz factor $\gamma=1/\sqrt{1-v^{2}/c^{2}}=1.42$) in the ion storage ring (ESR) were observed to decay by electron capture

\begin{equation}\label{decay}
^{142}_{61}Pm^{60+} + e^{-} \rightarrow ^{142}_{60}Nd^{60+} + \nu_{e}
\end{equation}

leaving a bare $^{142}_{60}Nd^{60+}$ nucleus and a neutrino in a two-body final state.

It was found that in such experiments the decay rate $R(t)$ - in addition to the exponential law - exhibits an oscillatory time modulation. The decay can be described by \cite{Kienle2013}

\begin{equation}\label{decay rate}
R(t) \propto e^{-\lambda t} (1+a\cdot\cos(\omega t +\phi)).
\end{equation}

The best-fit values \cite{Kienle2013} were $\lambda = 0.0130(8)$ s$^{-1}$, $\omega = 0.884(14)$ s$^{-1}$ (period $T_{osc}=7.11(11)$ s), amplitude $a=0.107(24)$ and phase $\phi=2.35(48)$ rad.

In the literature the question has been discussed if this oscillatory modulation is connected to the mass eigenstates of the emitted neutrino. A fundamental theory is still not available. However, it has been argued that the final state after electron capture is a superposition of two channels, which correspond to the two (neglecting the third one) mass eigenstates ($m_{1},m_{2}$) of the neutrino, resulting in an oscillatory frequency \cite{Kienle2013},\cite{Ivanov2009}

\begin{equation}\label{omega21}
\omega_{21}=\Delta m^{2}_{21} c^{2}/ (2M_{p}\hbar).
\end{equation}

Here, $\Delta m^{2}_{21} = m^{2}_{2} - m^{2}_{1}$ is the mass-squared difference of the two mass-eigenstate neutrinos \cite{Beringer2012}, $M_{p}$ is the mass of the decaying parent nucleus, and $\hbar$ and $c$ are Planck's constant divided by $2\pi$ and the speed of light, respectively. Using $\omega_{21} = \omega$ of ref. \cite{Kienle2013} and considering the relativistic time dilatation it has been suggested:

\begin{equation}\label{msquared}
\Delta m_{21}^{2}c^{4} = 2\hbar\omega\gamma M_{p} c^{2}
\end{equation}
However, eq. (\ref{msquared}) gives a value $\Delta m^{2}_{21}=2.19(3)\cdot 10^{-4}$ eV$^2$/c$^{4}$, which is nearly three times larger than $\Delta m^{2}_{21}=(0.754^{+0.026}_{-0.022})\cdot 10^{-4}$eV$^2$/c$^{4}$ determined in a global fit of the results obtained in reactor and solar neutrino experiments \cite{Fogli2012}.

There has been a long controversy in the literature doubting the validity of eq. (\ref{omega21}) on the basis of quantum mechanics and, in particular, questioning the presence of such interference effects in the decay rate observed in electron-capture storage ring experiments (see, e.g., \cite{Ivanov2009}, \cite{Cohen2009} - \cite{Giunti2009}). However, more recently, it was emphasized \cite{Gal2014}, \cite{Peshkin2014} that such oscillations in the EC decay rate might indeed be caused by interference effects due to the weak interaction \cite{Peshkin2014} or entanglement \cite{Gal2014}. For example, this is clearly stated in ref. \cite{Peshkin2014}: "Quantum mechanics permits oscillations in the rate of decay by electron capture to arise from interference effects proportional to the neutrino mass difference. That comes about because of indirect coupling of two neutrino mass channels through their direct coupling to the decaying ion by the weak interaction." Unfortunately, at present a theoretical foundation of eq. (\ref{omega21}) is still unclear.

In the following we will not further discuss this issue, but assume that such an indirect interaction (resulting in eq. (\ref{omega21})) does indeed occur and concentrate on the time-dilatation transformations according to Relativity Theory. After applying these transformations consistently (see sections 3 and 4), we find a new relation (eq. (\ref{msquarednew}) instead of eq. (\ref{msquared})) according to which the value for the neutrino mass-squared difference $\Delta m^{2}_{21}$ derived from the EC storage-ring experiment (see section 4) is in full agreement with $\Delta m^{2}_{21}$ determined by reactor and solar neutrino measurements. To further examine this new relation between EC-decay oscillations and $\Delta m^{2}_{21}$ additional experiments with different orbital speed, i.e. different $\gamma$ factor, should be performed.

\subsection{2. Time dilatation in accelerated systems}

Time dilatation occurs in systems with quasi-circular motion like in experiments using storage rings or high-speed centrifuges (rotors) \cite{WPotzel}, where accelerations in Minkowski space-time are present \cite{Ryder}. Such accelerations are due to changing direction but at constant speed. 


In a circular motion with an orbit of radius $R$, orbital velocity $v$, orbital frequency $\Omega$ (e.g., in a rotor), the frequency due to time dilatation is given by \cite{Ryder}, \cite{Schutz}


\begin{equation}\label{frequency shift}
f'=f\sqrt{1-\Omega^{2}R^{2}/c^{2}}=f\sqrt{1-v^{2}/c^{2}}=f/\gamma.
\end{equation}




\subsection{3. Time dilatation between the neutrino system and the parent system in EC storage-ring experiments}

When the parent ion (P) coasting in the storage ring decays to the daughter ion (D) and a neutrino, the daughter ion is now coasting and being accelerated, like the parent ion before the decay. However, the neutrino $\nu$ being a neutral particle is moving on a rectilinear path and is \textit{not} accelerated. As a consequence, the clock in the $\nu$ system (frame) runs faster than the clock in the parent (daughter) frame. This effect due to different accelerations in the $\nu$ and daughter frames is required by Relativity Theory, but has not been considered in the past.

Following Ref.\cite{Peshkin2014}, oscillations in the EC-decay rate may be caused by indirect coupling of two neutrino mass channels which are directly coupled to the decaying ion by the weak interaction. If information is transferred - e.g., by the weak interaction \cite{Peshkin2014} or via entanglement \cite{Gal2014} - from the $\nu$ frame to the daughter in the P frame, this time dilatation due to different acceleration has to be taken into account. As a consequence of this information transfer, the energy splitting $(E_{\nu2}-E_{\nu1})$ in the $\nu$ system as recorded by the daughter in the P frame is determined by

\begin{equation}\label{energy in P}
\Delta E^{(P)}_{D21}=E^{(P)}_{D2}-E^{(P)}_{D1}=\gamma (E_{\nu2}-E_{\nu1}),
\end{equation}

where $E^{(P)}_{D1}$ and $E^{(P)}_{D2}$ are the energy states of the daughter in the P system.
Eq. (\ref{energy in P}) gives rise to the interference term in eq. (\ref{decay rate}). Neglecting the phase $\phi$ we have \cite{Gal2014}

\begin{eqnarray}\label{interference term}
cos(\frac{\Delta E^{(P)}_{D21}}{\hbar}\cdot t_{P})=cos[\frac{\gamma (E_{\nu2}-E_{\nu1})}{\hbar}\cdot t_{P}]=
\nonumber\\
=cos(\frac{\gamma \Delta m^{2}_{21}c^{4}}{2\hbar M_{p}c^{2}}t_{p})
\end{eqnarray}
 
which we still have to transform to the laboratory frame.

\subsection{4. Time transformation between the daughter and the laboratory frames}

Transforming further to the laboratory frame we have \cite{Gal2014}

\begin{equation}
\Delta E^{(L)}_{21}=\gamma \Delta E^{(P)}_{D21}=\frac{\gamma^{2}\Delta m^{2}_{21}c^{4}}{2M_{p}c^{2}}
\end{equation}

This leads to a frequency of daughter-ion oscillations in the laboratory frame L:

\begin{equation}\label{omega}
\omega_{L}=\frac{\gamma^{2}\Delta m^{2}_{21}c^{4}}{2\hbar M_{p}c^{2}}
\end{equation}

or

\begin{equation}\label{msquarednew}
\Delta m^{2}_{21}c^{4}=\frac{2\hbar \omega_{L}M_{p}c^{2}}{\gamma^{2}}
\end{equation}

Using $\omega_{L}=\omega=0.884(14)$s$^{-1}$ and $\gamma=1.42$ \cite{Kienle2013} we get $\Delta m^{2}_{21}c^{4}=(0.768\pm 0.012)\cdot 10^{-4}$eV$^{2}$ which is in full agreement with $\Delta m^{2}_{21}c^{4}=(0.754^{+0.026}_{-0.022})\cdot 10^{-4}$eV$^{2}$ obtained from a global analysis of neutrino masses and mixing \cite{Fogli2012}.

We want to emphasize that in any storage-ring experiments where information is transferred between charged and neutral particles, time dilatation in accelerated systems has to be considered. However, if the $\nu$-oscillations would directly be observed in the $\nu$-system (which has not been done and would be very difficult to do) the transformation from the $\nu$-frame to the laboratory frame L would just involve the factor $\gamma$, since - due to time dilatation - the oscillation frequency in the $\nu$-frame is reduced by $\gamma$ compared to the L-frame.

To check eq. (\ref{omega}) experimentally, the orbital speed $v$, i.e., the value for $\gamma$ could be changed in future experiments. According to eq. (\ref{omega}), $\omega_{L}$ is expected to show a $\gamma^{2}$ dependence and \textit{not} a linear dependence as suggested in \cite{Gal2014}. An observation of a $\gamma^{2}$ dependence would also advocate eq. (\ref{omega21}). Our eq. (\ref{msquarednew}) replaces eq. (\ref{msquared}). The main new aspect is the time dilatation between the $\nu$ system and the parent frame due to different accelerations (described in section 3) which is then followed (in section 4) by a transformation from the daughter to the laboratory frames. The $(1/M_{P})$-dependence of $\omega_{L}$ in eq. (\ref{omega}) has already been confirmed by experiments where in addition to the $^{142}Pm$ and $^{140}Pr$ isotopes also $^{122}I$-ions were investigated \cite{Kienle2009}.

\subsection{5. Conclusions}

The results of electron-capture storage-ring experiments in connection with neutrino oscillations have been re-evaluated assuming that information is transferred from the $\nu$ frame to the P(D) frame, e.g., by entanglement or weak interaction as suggested in Refs. \cite{Gal2014} and \cite{Peshkin2014}, respectively. After the decay, the neutrino - being an electrically uncharged particle - is \textit{not} accelerated, but moving on a rectilinear path, in contrast to the daughter ion. As a consequence, the clock in the $\nu$ frame runs faster than the clock in the daughter frame. Taking this aspect into account, we derive a value $\Delta m^{2}_{21}$ from the observed EC-decay oscillations which fully agrees with that obtained from other neutrino-oscillation measurements. Thus it appears that EC storage-ring experiments could turn out to be an accurate method for the determination of neutrino mass-squared differences. The $\gamma^2$ dependence should be investigated by repeating the experiment with a few different $\gamma$ values. Those results would indirectly also test the validity of eq. (\ref{omega21}). Unfortunately, a fundamental theory which describes this information transfer and thus relates the EC-decay oscillations to the neutrino mass-squared difference is still missing.

\begin{acknowledgments}
It is a great pleasure to thank Alejandro Ibarra, Thomas Faestermann, Alexander Merle, Avraham Gal, and Murray Peshkin for stimulating discussions.
\end{acknowledgments}


\begin{thebibliography}{99}

\bibitem{Litvinov2008} Yu.A. Litvinov et al., Phys. Lett. B\textbf{664}, 162 (2008).
\bibitem{Kienle2013} P. Kienle et al., Phys. Lett. B\textbf{726}, 638 (2013) and references therein.
\bibitem{Ivanov2009} A.N. Ivanov and P. Kienle, Phys. Rev. Lett. \textbf{103}, 062502 (2009).
\bibitem{Beringer2012} J. Beringer et al. (Particle Data Group), Phys. Rev. D \textbf{86}, 010001 (2012).
\bibitem{Fogli2012} G.L. Fogli et al., Phys. Rev. D \textbf{86}, 013012 (2012).

\bibitem{Cohen2009} A.G. Cohen, S.L. Glashow, and Z. Ligeti, Phys. Lett. B \textbf{678}, 191 (2009).
\bibitem{Flambaum2010} V.V. Flambaum, Phys. Rev. Lett. \textbf{104}, 159201 (2010).
\bibitem{Ivanov2010} A.N. Ivanov and P. Kienle, Phys. Rev. Lett. \textbf{104}, 159202 (2010).
\bibitem{Lipkin2010} H.J. Lipkin, Phys. Rev. Lett. \textbf{104}, 159203 (2010) and arXiv: 0910.5049 (2009).
\bibitem{Merle2009} A. Merle, Phys. Rev. C \textbf{80}, 054616 (2009).
\bibitem{MerlearXiv} A. Merle, arXiv: 1004.2347 (2010).
\bibitem{Kienert} H. Kienert et al., arXiv: 0808.2389 (2008).
\bibitem{Gal2010} A. Gal, Nucl. Phys. A \textbf{842}, 102 (2010).

\bibitem{Giunti2008} C. Giunti, Phys. Lett. \textbf{B665}, 92 (2008).
\bibitem{Giunti2009} C. Giunti, Nucl. Phys. Suppl. \textbf{188}, 43 (2009).

\bibitem{Gal2014} A. Gal, arXiv: 1407.1789v4 (2014).
\bibitem{Peshkin2014} M. Peshkin, arXiv: 1409.0037 (2014).
\bibitem{WPotzel} For a review, see W. Potzel in \textit{The Rudolf M\"ossbauer Story}, M. Kalvius and P. Kienle eds., chapter 14: \textit{Relativistic Phenomena Investigated by the M\"ossbauer Effect}, (Heidelberg: Springer-Verlag, 2012), pp.263, and references therein.


\bibitem{Ryder} L. Ryder, \textit{Introduction to General Relativity}, chapter 2.3 Rotating frames: the Sagnac effect (Cambridge University Press, Cambridge, 2009).
\bibitem{Schutz} B. Schutz, \textit{A first course in General Relativity}, second edition (Cambridge University Press, Cambridge, 2009). 

\bibitem{Kienle2009} P. Kienle, Nucl. Phys. A \textbf{827}, 510c (2009).






















\end{thebibliography}

\end{document}